\documentstyle[12pt]{article}
\normalbaselines
\begin{document}
\newcommand{\be}{\begin{equation}}
\newcommand{\ee}{\end{equation}}        
\newcommand{\w}{wavelet}
\newcommand{\an}{analysis}
\newcommand{\1}{one-dimensional}
\newcommand{\2}{two-dimensional}
\newcommand{\m}{two-microlocal}
\newcommand{\co}{coefficient}
\newcommand{\scl}{\varphi}
\newcommand{\mr}{multiresolution}
\bibliographystyle{unstr}
\vbox{\vspace{6mm}}
\begin{center}
{\large \bf SYMMETRIES AND WAVELETS \\  [7mm]}
I.M. Dremin      \\
{\it Lebedev Physical Institute, 119991 Moscow, Russia}   \\
\end{center}
\vspace{2mm}
\begin{abstract}
Wavelet analysis of different patterns reveals some symmetries and
singularities otherwise hidden in the pattern. Its general methods
are briefly reviewed. Examples from turbulence, cavitation, Cherenkov
gluon emission and quark-gluon jets structure in QCD are considered.
\end{abstract}
Symmetries are crucial in the surrounding us world and therefore lie
at the very heart of any theoretical construction. Symmetries can be either
global or local. The invariance under the global gauge transformation
requires a definite symmetry between particles and antiparticles. More
important consequences appear from the requirement of the local gauge
invariance. Both in QED and QCD it defines the interaction lagrangian and,
consequently, the symmetries of main equations. In classical and quantum physics
the symmetries of equations and of initial (boundary) conditions determine
the symmetries of the corresponding patterns observed. Thus the problem of
the local analysis of these patterns should be solved.

One of the most popular methods is to use the Fourier transform. The set of
functions (sine, cosine and imaginary exponents) is such that all of them are
widespread along their definition axis. Thus this transform is a global one.
With its help one can get the knowledge of frequencies (scales) important in
the problem considered but can not learn about their locations. In other
words, this transform is suited for stationary (homogeneous) processes. For
non-stationary (inhomogeneous) processes one should apply a local transform.
With the above set of functions, it becomes possible only if one uses the
so-called windowed Fourier transform. Namely, the analyzed function is
multiplied by an auxiliary function steeply decreasing (or equal to
zero) outside some finite range (window) of its variables. Sometimes, this leads
to uncontrollable consequences.

The only way to control such transform is to build up a complete set of
orthogonal functions with the compact support. This program is realized with
the help of wavelets \cite{meye, daub, dine}. They allow for the direct and
inverse local transform to be done. Thus one gets the local scale (frequency)
characteristics (fluctuations) of the analyzed process within any required
resolution. Moreover, this resolution is self-adjustable in contrast to fixed
windows in the Fourier transform. It means that wavelets automatically 
construct the so-called Heisenberg windows adjusted to the local properties of
the considered function. To resolve its fast variation with high frequencies
they admit small time-window but large frequency-window while for low
frequencies the wide time-window and narrow frequency-window are used,
i.e., the width of the window is proportional to its mean value.
That is why they are called the relative band-width filters.

Wavelets originate from the functional equation which relates the so-called
scaling function with its shifted and translated version. The coefficients $h_k$
of this linear equation define in a unique way the explicit form of the
wavelet which is not used directly, however. The wavelet transform coefficients
can be explicitly obtained from the iterative equations containing $h_k$. This 
procedure is called the fast wavelet transform. It allows for the complete 
decomposition of the considered process at any resolution level and does not
require any integrartion. Computer calculations are done quickly.

For those dealing with quantum field theory, I'd like to mention also that
\w\ \an\ actually is very close to the renormalization group approach (see,
e.g., \cite{fesh}) because it uses the translated versions of the same
function. These features must be further exploited.

The equation for the scaling function $\varphi (x)$ is
\be
\scl (x)=\sqrt 2 \sum _{k=0}^{2M-1}h_k\scl (2x-k)    \label{sclx}
\ee
with the dyadic dilation 2, integer translation $k$ and the coefficients
$h_k$ determined from conditions of orthogonality (both mutual and to
some polynomials) after one chooses their number $2M$.
If the scaling function is known, one can form a
"mother \w\ " (or a basic \w\ ) $\psi (x)$ according to
\be
\psi (x)=\sqrt 2 \sum _{k=0}^{2M-1}g_k\scl (2x-k),  \label{psix}
\ee
where
\be
g_k=(-1)^{k}h_{2M-k-1}.   \label{gkhk}
\ee

The dilated and translated versions of the scaling function $\scl $ and
the "mother \w " $\psi $
\be
\scl _{j,k}=2^{j/2}\scl (2^jx-k),    \label{sclj}
\ee
\be
\psi _{j,k}=2^{j/2}\psi (2^jx-k)     \label{psij}
\ee
form the orthonormal basis.
 
One may decompose any function $f$ of $L^2(R)$ at any resolution level $j_n$
in a series
\be
f=\sum _{k}s_{j_n,k}\scl _{j_n,k}+\sum _{j\geq j_n,k}d_{j,k}\psi _{j,k}.  \label{fdec}
\ee
The \w\ \co s $s_{j,k}$ and $d_{j,k}$ can be calculated as
\be
s_{j,k}=\int dxf(x)\scl _{j,k}(x),   \label{ssjk}
\ee
\be
d_{j,k}=\int dxf(x)\psi _{j,k}(x).   \label{ddjk}
\ee
However, in practice their values are determined from the fast \w\ transform.
In general, one can get the iterative formulas of the fast \w\ transform
\be
s_{j+1,k}=\sum _mh_ms_{j,2k+m},      \label{sshs}
\ee
\be
d_{j+1,k}=\sum _mg_ms_{j,2k+m}      \label{dsgs}
\ee
where
\be
s_{0,k}=\int dxf(x)\scl (x-k).         \label{sifs}
\ee
These equations yield fast algorithms (the so-called pyramid algorithms) for
computing the \w\ \co s, asking now just for $O(N)$ operations to be done.
Starting from $s_{0,k}$, one computes all other \co s provided the \co s 
$h_m,\; g_m$ are known. The explicit shape of the \w\ is not used.
The \w\ \co s show the strength of fluctuations of $f$ at the location $k$
and the scale $j$.

The behavoir of \w\ \co s is closely related to the structure of the local
singularities of the function $f$ and therefore can provide the knowledge
which class of functions it belongs to. Also, the fractal properties of the
analyzed set can be determined from the partition function $Z_q(j)$.
Namely, let us consider the sum $Z_q$ of the $q$-th moments
of the coefficients of the \w\ transform at various scales $j$
\be
Z_q(j)=\sum _k\vert d_{j,k}\vert ^q ,    \label{zq}
\ee
where the sum is over the maxima of $\vert d_{j,k}\vert $. Then it was shown
that for a fractal signal this sum should behave as
\be
Z_q(j)\propto 2^{j[\tau (q) +\frac {q}{2}]} , \label{zqj}
\ee
i.e.,
\be
\log Z_q(j)\propto j[\tau (q)+\frac {q}{2}].   \label{lzq}
\ee
Thus the necessary condition for a signal to possess fractal properties is the
linear dependence of $\log Z_q(j)$ on the level number $j$. If this requirement
is fulfilled the dependence of $\tau $ on $q$ shows whether the signal is
monofractal or multifractal. Monofractal signals are characterized by a single
dimension and, therefore, by a linear dependence of $\tau $ on $q$, whereas
multifractal ones are described by a set of such dimensions, i.e., by non-linear
functions $\tau (q)$. Monofractal signals are homogeneous, in the sense that 
they have the same scaling properties throughout the entire signal. Multifractal
signals, on the other hand, can be decomposed into many subsets characterized by 
different local dimensions, quantified by a weight function. The \w\ transform
removes lowest polynomial trends that could cause the traditional box-counting
techniques to fail in quantifying the local scaling of the signal. The function 
$\tau (q)$ can be considered as a scale-independent measure of the fractal 
signal. It can be further related to the Renyi dimensions, Hurst and H{\"o}lder
exponents (for more detail, see Ref. \cite{dwdk}).

The \w\ transform can be also used for representation of differential operators,
for solving the differential equations etc.

Let me describe the \w\ application to the \an\ of some symmetrical patterns.
Many such patterns are observed, e.g., on the water surface \cite{patt}
(see Figs. 1, 2). Macroscopically they are described by some order parameters.
The \w\ decomposition of the observed pattern allows to ascribe the definite
scale at any location within the pattern and extract wave vector information
from the pattern.
In particular, the regions with scales within the chosen
range can be separated. E.g., the stability regions with large
scales are shown in Fig. 3 in white. Thus the transition from microscopic
pattern to its macroscopic description becomes feasible. The location of
some defects (broken symmetry) is easily recognized as well.

Another interesting effect is the turbulence. Its \w\ \an\ reveals the 
fractal properties of the velocity field. Closer to the topic of this
conference are the similar patterns observed in jet production in
$e^{+}e^{-}$-annihilation at high energies (for the review see
\cite{dwdk, dgar}). The partonic cascade structure 
with "jets inside jets inside jets" leads to the mono- and multifractal
distribution of the created particles within the available phase space
(see \cite{dwdk}). The fluctuations increase in smaller phase space bins,
and, correspondingly, the factorial moments of the multiplicity distributions
should increase linearly (for monofractals) for smaller bins and flatten off
for multifractals. This is demonstrated in Fig. 4.
It is explained in QCD \cite{38, 68, 70} as a consequence of the asymptotic
freedom behavior of the coupling strength.

The fractal properties have been also discovered by the \w\ \an\ in the 
distributions of the heartbeat intervals. One would be inclined to consider
this process as a completely periodic stationary one. In fact, it has been found
that it possesses the fractal properties and the distribution moments
(correlations) at different scales behave in a different way for healthy and
diseased patients \cite{tfte, agis}. The results of the \w\ \an\ were proposed
as a first clinically significant measure of heart disease.

The functions somewhat similar, at first sight, to those for heartbeat intervals
are obtained for the pressure variation in gas turbine compressors. However 
their \w\ \an\ revealed absence of fractal properties with some signature of a
singular behavior. More important discovery is that the behavior of the
dispersion of the \w\ \co s at some resolution level can be used as a precursor
of the extremely dangerous effect \cite{dfin}. It is initiated by some instability
(singularity), reminds the cavitation and is called the stall. This instability 
results in the complete damage of the engine and an aircraft crash.

Coming back to particle interactions, let me mention the important field of \w\
application to pattern recognition in multiparticle production. One can ask
what kind of patterns are formed within the available phase space by particles
created in very high energy interactions. This problem becomes experimentally
feasible, e.g., in $AA$-collisions at RHIC where more than 4000 charged
particles are on average produced in a single event at c.m. energy 130 GeV/c.
In a search for the ring-like structure implied by ideas on Cherenkov gluon
radiation \cite{drem}, the \w\ \an\ of several events of central Pb-Pb 
collisions at l.s. energy 158 GeV/c with more than 1000 charged secondaries
was done. Some events possessing such a symmetry in the long-range correlations
have been found \cite{dikk} (see Fig. 4). This \an\ can be extended to search
for other symmetry patterns as well.

In conclusion, I'd like to stress how powerful is the method of the \w\ \an\
to search for symmetries as is seen already from the above examples. However,
its use is at the very beginning now and further results are coming.\\

{\bf Figure captions}\\

Fig. 1 The pattern in experimental convection (from \cite{aste}).\\

Fig. 2 The symmetry and imperfections in a hexagonal Benard convection cell
(from \cite{kosc}).\\

Fig. 3 The original pattern (b) is \w\ analyzed and (a) the regions of high
wave vectors ($k>k_B$) are marked in white, (c) the wave-number histogram
computed with
the \w\ algorithm (solid line) is more precise than that with the Fourier
transform (dotted line); $k_B$ is the wave number separating regions with
straight rolls stable ($k>k_B$) and unstable to bending, (d) the
correlations of the pattern along various cross sections
are shown to be similar (from \cite{patt}).\\

Fig. 4 The factorial moments $F_q$ of multiplicity distributions in
$e^{+}e^{-}$-collisions at $Z^0$ energy show the mono- and multifractal
behavior (linear and
curved parts of the lines, correspondingly) of particles pattern in the
available phase space (from \cite{opal}). Increasing $z$ corresponds to
smaller bins. Experimental data are shown by dots.
Different lines correspond to different calculations within some
approximations of QCD.\\

Fig. 5 The ring-like structure of some events of Pb-Pb interaction at 158 GeV/c
(from \cite{dikk}) revealed by the \w\ \an\ demonstrates the symmetry otherwise
hidden in a huge background. 


\begin{thebibliography}{99}
\bibitem{meye}
Meyer Y {\it Wavelets: Algorithms and Applications} (Philadelphia: SIAM, 1993)
\bibitem{daub}
Daubechies I {\it Ten Lectures on Wavelets} (Philadelphia: SIAM, 1991)
\bibitem{dine}
Dremin I M, Ivanov A V and Nechitailo V A {\it UFN} {\bf 171} 465 (2001);
{\it Physics-Uspekhi} {\bf 44} (May 2001)
\bibitem{fesh}
Feshbach H {\it MIT-CTP-2192} (1993)
\bibitem{dwdk}
De Wolf E, Dremin I M and Kittel W {\it Phys Rep} {\bf 270} 1 (1996) 
\bibitem{patt}
Bowman C and Newell A C {\it Rev Mod Phys} {\bf 70} 289 (1998)
\bibitem{dgar}
Dremin I M and Gary J W {\it Phys Rep} {\bf 349} 301 (2001)
\bibitem{38}
Dokshitzer Yu L and Dremin I M {\it Nucl Phys} {\bf B 402} 139 (1993) 
\bibitem{68}
Ochs W and Wosiek J {\it Phys Lett} {\bf B 289} 159 (1992); {\bf B 304} 144 (1993)
\bibitem{70}
Brax Ph, Meunier J L and Peschanski R {\it Z Phys} {\bf C 62} 649 (1994)
\bibitem{tfte}
Thurner S, Feurstein M C and Teich M C {\it Phys Rev Lett} {\bf 80} 1544 (1998)
\bibitem{agis}
Amaral L A N, Goldberger A L, Ivanov P C et al {\it Phys Rev Lett} {\bf 81}
2388 (1998)
\bibitem{dfin}
Dremin I M, Furletov V I, Ivanov O V et al {\it Control Engineering Practice}
(2001) (to be published)
\bibitem{drem}
Dremin I M {\it JETP Lett} {\bf 30} 140 (1979); {\it Yad Fiz} {\bf 33} 1357 (1981)
\bibitem{dikk}
Dremin I M, Ivanov O V, Kalinin S A et al {\it Phys Lett} {\bf B 499} 97 (2001)
\bibitem{aste}
Assenheimer M and Steinberg V {\it Nature} {\bf 367} 345 (1994)
\bibitem{kosc}
Koschmieder E L {\it Adv Chem Phys} {\bf 26} 177 (1974)
\bibitem{opal}
L3 Collaboration, Acciari M et al {\it Phys Lett} {\bf B 428} 186 (1998)
\end{thebibliography}
\end{document}